\documentclass[acmsmall,screen]{acmart}
\acmConference[SE 2030]{International Workshop on Software Engineering in 2030}{November 2024}{Puerto Galinàs (Brazil)}

\AtBeginDocument{%
  }

\setcopyright{acmlicensed}
\copyrightyear{2024}
\acmYear{2024}
\acmDOI{XXXXXXX.XXXXXXX}

\acmJournal{TOSEM}
\acmVolume{1}
\acmNumber{1}
\acmArticle{1}
\acmMonth{11}

\begin{document}

\title{Morescient GAI for Software Engineering (Extended Version)}


\author{Marcus Kessel}
\email{marcus.kessel@uni-mannheim.de}
\author{Colin Atkinson}
\email{colin.atkinson@uni-mannheim.de}
\affiliation{%
  \institution{University of Mannheim}
  \city{Mannheim}
  \country{Germany}}


\begin{abstract}
The ability of Generative AI (GAI) technology to automatically check, synthesize and modify software engineering artifacts promises to revolutionize all aspects of software engineering. Using GAI for software engineering tasks is consequently one of the most rapidly expanding fields of software engineering research, with over a hundred LLM-based code models having been published since 2021. However, the overwhelming majority of existing code models share a major weakness -- they are exclusively trained on the syntactic facet of software, significantly lowering their trustworthiness in tasks dependent on software semantics. To address this problem, a new class of ``Morescient'' GAI is needed that is ``aware'' of (i.e., trained on) both the semantic and static facets of software. This, in turn, will require a new generation of software observation platforms capable of generating large quantities of execution observations in a structured and readily analyzable way. In this paper, we present a vision and roadmap for how such ``Morescient'' GAI models can be engineered, evolved and disseminated according to the principles of open science.
\end{abstract}





\maketitle



\section{Introduction}
\label{sec:introduction}

Generative-AI-based tools, and particularly Large Language Models (LLMs) of the kind made famous by OpenAI's ChatGPT, are set to become one of the most disruptive technologies in software engineering for decades. They are not only able to synthesize software artifacts (including source code) from simple, natural language prompts \cite{chen2021evaluating}, they can automatically perform many other software engineering tasks such as repairing, refactoring and explaining code, generating test cases and comprehending programs. Training and using LLMs for software engineering tasks, which Hou et al. \cite{hou2024large} call \textit{LLM4SE} and Lo calls \textit{AI4SE} \cite{10449668}, has therefore rapidly become one of the largest research fields in software engineering with well over $100$ different models for code synthesis having already been published and/or described in the literature \cite{liu2023is,evalPlusLeaderboard}.

Despite their huge potential, however, most LLM4SE applications share one major Achilles' heel that significantly reduces their potential to improve productivity in mainstream software engineering projects -- they are exclusively trained on the \textit{syntactic facet} of software. Static properties of code are not only the most tangible manifestation of software, they are the most amenable to the traditional, static analysis approaches of computer science and the statistical NLP techniques of data science (cf. the naturalness of code principle \cite{10.1145/3212695}). However, code has another equally important facet, its \textit{run-time behavior} (a.k.a. as its dynamic semantics), which is much less tangible and cannot, in general, be deduced analytically from the syntactic facet (cf. Rice's theorem \cite{riceTheorem}). As a result, the ``true'' or ``de facto'' dynamic semantics of a non-trivial piece of code can only be determined by observing its behavior in response to known stimuli under known conditions (i.e., through software testing) \cite{6963470,ammann2016introduction}.

Since the vast majority of code LLMs created to date are exclusively trained on the syntactic facet of software, they have no direct knowledge of the semantic facet, and are thus ``unaware'' of the actual behavior of software. Thus, when an LLM exclusively trained on syntactic data is asked to create a procedure to ``sort'' an array of elements, for example, it can only use the linguistic (i.e., identifier-based) information it has collected syntactically about procedures  related to ``sort'' rather than procedures that demonstrably deliver ``sort'' behavior at run-time, because it has never been exposed to data that captures the ``de facto'' behavior of such procedures. In addition to their ignorance of true functional properties, LLMs are also unaware of another critical aspect of software: non-functional properties such as performance (e.g., speed and resource utilization) which can only be obtained at run-time in a target execution environment. Functional and non-functional properties are crucial aspects of software development that play a significant role in ensuring the quality, reliability, and overall success of software systems \cite{basili1994goal}. Neglecting any of them can have significant consequences such as higher risks and operational costs (e.g., blindly choosing a generated code artifact that turns out to have poor resource usage behavior in a cloud-based environment).

This ignorance of de facto behavior directly contributes to the low trustworthiness of the current generation of code LLMs. For example, in a recent experiment (a replication) \cite{multiple} we showed that code synthesis solutions for Java coding problems returned by OpenAI's \textsc{Davinci} model are, on average, correct only $42\%$ of the time \cite{KESSEL2024111971}, and the solutions returned by two other popular models, \textsc{Codegen} and \textsc{InCoder}, are correct only $22\%$ and $9\%$ of the time, respectively. On the other hand, although more recent studies have shown substantial gains on widely used, Python-based benchmarks like \textsc{HumanEval} and \textsc{MBPP} and their improved derivatives\footnote{Even the benchmarks were criticized for missing true behavior awareness, since they contained weak tests} like \textsc{EvalPlus} \cite{liu2023is}, these advances come with a major caveat: the benchmarks are becoming increasingly saturated. In fact, some models are already converging to optimal precision, reducing their utility for evaluating model performance for practitioners. 

These problems have contributed to growing skepticism about the representativeness of popular benchmarks, and the potential for data leakage \cite{matton2024leakagecodegenerationevaluation} (i.e., models being trained on solutions to the code challenges used to evaluate them). As a result, there is an increasing concern that current benchmarks may not accurately reflect likely performance in real-world coding tasks, but merely indicate their performance on artificially-simple coding problems. Researchers and practitioners alike are therefore increasingly skeptical about the reliability of AI-generated code, with major tech companies like Microsoft, Google and Meta reporting that their developers typically reject around two-thirds of AI-powered code recommendations (e.g., GitHub Copilot \cite{copilotReport2024}). This widespread untrustworthiness of GAI raises concerns that potential productivity gains from automated code synthesis may be offset by the need for additional verification and validation effort to ensure the quality of the resulting software system \cite{ammann2016introduction}.

To address the lack of awareness of de facto run-time behavior, we believe a completely new class of code LLMs is required that ``knows about'' the true dynamic semantics of code, alongside the usual syntactic properties, because they have been trained on both facets of software rather just the latter. We characterize such LLMs as ``morescient'' -- a neologism derived from the Latin words ``mores'' (behavior) and ``scire'' (to know). Our underlying hypothesis is that morescient LLMs will deliver much more trustworthy results \cite{trustworthy_10.1145/3555803,10449668} than traditional trained LLMs, and therefore will be much more useful in practical software engineering projects\cite{sun2024trustllm}.

The outline of the paper is as follows. First, in Section \ref{sec:limitations}, we discuss the current state of the art in morescient GAI and the main obstacles to achieving it. Next, in Section \ref{sec:stimulus_response_data_structures}, we present the types of data structures and platforms necessary for realizing morescient GAI. This is followed by Section \ref{sec:continual_srh}, where we outline our vision for an open, continual approach to maintaining an evolving dataset of observations of the same order of magnitude as current syntactic code datasets. Finally, in Section \ref{sec:usage}, we provide a roadmap for integrating true-behavior awareness into generative AI for software engineering. We conclude this work with some closing remarks in Section \ref{sec:conclusion}.
\section{State of the Art}
\label{sec:limitations}

At the time of writing, the overwhelming majority of GAI code models have been trained on statically derived data and raw code. There has been a small number of attempts to include de facto (i.e., observed) run-time behavior in the training or use of code LLMs, but since they are highly limited in their scope and scale, we refer to them as semi-morescient. In the following subsection we describe these preliminary attempts at morescience, before summarizing the main obstacles that stand in the way of achieving this in a comprehensive and systematic way in the subsequent subsection.

\subsection{Semi-Morescient Approaches}

To date, researchers have explored three main lines of research to improve the morescience of GAI-derived code. The first line, which we refer to as ``morescient training'', aims to include observations of code modules' run-time behavior in the collection of data used to train and fine-tune code LLMs. The second line, which we refer to as ``morescient usage'', aims to improve the results gained from code LLM queries by selecting and refining results using feedback from the compilation and/or execution of returned candidates. The third line of research, which we refer to as ``morescient benchmarking'', uses execution-based feedback to improve the benchmarking of code LLMs. Each of these is elaborated further below.

\subsubsection*{Morescient Training}

This line of research leverages execution information in the initial training and fine-tuning of execution-aware code models. Notable examples include \textsc{CodeExecutor} \cite{liu-etal-2023-code} and \textsc{TRACED} \cite{traced_10.1145/3597503.3608140}. The latter, in particular, incorporates run-time trace information into its training data to enhance understanding of program behavior across multiple downstream tasks: semantic clone retrieval, vulnerability detection, and static execution estimation (e.g., predicting which paths or code elements will be covered).

More recent research has also attempted to train models to comprehend and reason about operational semantics in a step-by-step execution style, including \textsc{scratchpad} \cite{nye2021workscratchpadsintermediatecomputation}, \textsc{NeXt} \cite{ni2024nextteachinglargelanguage}, and \textsc{SemCoder} \cite{ding2024semcodertrainingcodelanguage}. While \textsc{scratchpad} and \textsc{NeXt} employ trace reasoning formats of different kinds, \textsc{SemCoder} utilizes a monologue reasoning technique, where each execution step is verbalized to provide a human-readable representation of the code's execution.
At the model construction stage, learning-based techniques can be employed to improve the quality of the code LLMs generate for certain software engineering tasks. By incorporating behavioral feedback obtained from compilers and/or tests \cite{shojaee2023executionbased,liu2023rltf} within reinforcement learning frameworks, it is possible to identify functionally incorrect generated code and take corrective action.

GAI-based oracle recommendation approaches like \textsc{TOGA} \cite{10.1145/3510003.3510141} or program repair approaches \cite{10172803} are specifically trained using test code (e.g., harvested or generated unit test methods). \textsc{TOGA}, for instance, predicts oracle values by generating assertion statements for a provided test. Since test code embodies abstract semantics (i.e., high-level understanding of code module properties), such as inputs and outputs, they have elements of morescience. However, they can only be classified as semi-morescient because their ability to generate oracle values is critically dependent on accurate predictions of the code that is supposed to deliver the described behavior, including the prediction of the actual ``focal method'' -- that is, the portion of the code that is being tested by a test.

\subsubsection*{Morescient Usage}

This line of work uses execution information gathered through testing to select the best solution from multiple code generations. This can be accomplished by sampling from a set of potential solutions, such as those provided by \textsc{CodeT} \cite{chen2022codetcodegenerationgenerated} or \textsc{LEVER} \cite{ni2023lever} and selecting the one that reveals interesting discrepancies \cite{kesselGAI2024}. Some authors (e.g., \cite{chen2023teachinglargelanguagemodels}) apply this idea using the run-time feedback to iteratively improve the generated solution through so called ``self-debugging''.

The authors of \cite{li2024chaincodereasoninglanguage} further showcase the effectiveness of execution-based reasoning as a prompting technique by introducing the Chain of Code (CoC) approach, a novel prompting technique that generates code to reason through problems. By executing this generated code using either a code interpreter or a language model simulating the execution, CoC can tackle intricate coding problems.

\subsubsection*{Morescient Benchmarking}

Existing benchmarks that reason about the properties of LLM generated code have recently also become more aware of code behavior. While existing popular benchmarks like \textsc{HumanEval} and \textsc{MBPP} \cite{liu2023is}, or \textsc{CodeContests} \cite{doi:10.1126/science.abq1158}, include tests that describe the desired behavior in terms of abstract semantics (i.e., test inputs and expected outputs), they are only used to evaluate the performance of models and their code generation capabilities with respect to the property of functional correctness. However, they do not assess the other fundamental dimension of code models -- their ability to reason about the de facto behavior of code. \textsc{CRUXEval} \cite{gu2024cruxevalbenchmarkcodereasoning} is an ``execution benchmark'' for code reasoning, understanding, and execution evaluation that attempts to address this problem by using $800$ Python functions with corresponding input-output pairs to assess their input value prediction and output value prediction capabilities.

\subsection{Obstacles to Morescient GAI}

Although the aforementioned approaches have delivered useful insights how GAI technology can be made more aware of run-time behavior, they are all limited by a number of fundamental obstacles to the systematic, comprehensive and open development of morescient GAI, as summarized in the following sections.

\subsubsection*{Lack of Suitable Data Structures}

Current approaches are typically ad hoc solutions employing improvised data structures (and representations) as well as manually curated data sets, or data sets synthesized with the assistance of LLMs (e.g., \cite{gu2024cruxevalbenchmarkcodereasoning,ding2024semcodertrainingcodelanguage}). 
Although things have improved slightly, the availability of dynamic (observational) data remains significantly limited compared to syntactic data, hindering the development of fully morescient models. This is because real-world, high-quality corpora of run-time behavior (i.e., morescient data sets) are challenging to obtain and curate at a large scale \cite{10.1145/3236454.3236501,10.1145/3643742}. A prerequisite is parsable and executable code. However, even if code meets these criteria, observing its executable behavior poses a significant challenge due to the lack of documentation needed for testing  (i.e., lack of tests).

To achieve the desired scaling effects of LLMs for run-time behavior \cite{kaplan2020scalinglawsneurallanguage}, it is essential to develop more advanced LLM models supported by efficient data structures. These structures must be capable of storing or linking vast quantities of observation data -- both functional and non-functional -- and associating it with existing syntactic code models. Additionally, they must be designed with interoperability in mind, enabling seamless integration for the training and continuous improvement of LLMs.

The diversity of code semantics -- including behavioral and static -- can be captured by relationships over distinct dimensions. To effectively manage these relationships, they must be cleanly and explicitly stored. Consequently, the required data structures should support a unified framework that enables navigation over, and tracking of, these relationships (e.g., linking code modules to tests that executed them and to the observed behavior).

\subsubsection*{Fragmented and Poor Quality of Training Data}

The performance of LLMs is only as good as the quality of its training data, which depends on several factors, particularly when it comes to natural language text and code. It has been shown that LLMs replicate vulnerabilities, bugs, and poor code quality present in their training data \cite{chen2021evaluating,10.1145/3643674,asare_is_2023}. For example, while the \textsc{Codex} LLM powering GitHub Copilot can help prevent some types of ``simple, stupid'' bugs, it also generates code containing these bugs at twice the rate of correct code \cite{10174227}.

One essential factor affecting quality is deduplication of data and code, which is considered critical for the quality of training data sets \cite{10.1145/3359591.3359735,lee-etal-2022-deduplicating}.
Unfortunately, since there is no commonly accepted data structure to represent run-time behavior and its various representations in terms of dynamic semantics, there is a lack of understanding about what constitutes high-quality dynamic training data, beyond general guidelines learned from general-purpose, syntactic training data.

The current generation of code models suffers from poor data quality, primarily due to a critical oversight: unverified training code. Many data sets lack executable code, as checks for compilability (i.e., syntax and static semantics) and executability are often skipped because of the high costs and complexity involved (e.g., missing build information). Desirable properties like self-contained code modules (i.e., standalone modules that can run without external dependencies) are also difficult to identify automatically. As a result, there have been ad hoc attempts to manually curate or synthesize training data to bypass the challenges of collecting dynamic data, which has led to a fragmented data landscape. For example, Ding et al. found that roughly a quarter of the Python solutions in OSS-Instruct \cite{wei2024magicoderempoweringcodegeneration}, which were synthetically created from code LLMs, were not executable \cite{ding2024semcodertrainingcodelanguage}.

If this fundamental issue could be addressed, generated code would likely exhibit similar characteristics (i.e., executability), making large-scale data structures and platforms more feasible.

\subsubsection*{Lack of Structured Representations of Semantics}

The availability of specialized data structures for storing and linking semantic information represents just one aspect of the broader challenge of reasoning about functional and non-functional properties of software. Equally crucial is the consideration of how observed semantics are actually represented within these structures, effectively capturing their nuances and complexities.
The development of fully morescient GAI approaches is hindered by the lack of formal representations of run-time behavior, especially in how the true behavior of software is linked to the syntax in the data used to train LLMs. For operational semantics that describe actions in terms of step-by-step computations (e.g., debugging), effectively representing execution traces has been a persistent challenge in dynamic analysis. In the case of LLMs, even representing abstract semantics presents difficulties, such as selecting an appropriate encoding scheme to serialize input and output objects for abstract semantics (e.g., a suitable string representation for Java objects). Ideally, these semantic representations should be language-agnostic, enabling more efficient knowledge transfer across multiple programming languages. However, current approaches (e.g., \cite{traced_10.1145/3597503.3608140,ding2024semcodertrainingcodelanguage}) are limited, as they are typically designed for specific programming languages, restricting their applicability.

The choice of behavior representation has a substantial impact on how the model processes and generates results, including the textual output. Understanding how run-time behavior is represented in LLMs is essential for designing effective architectures, accurately evaluating performance metrics, and creating applications tailored to specific user needs based on perspectives on run-time behavior.

Developing a clear definition of software semantics is a challenging task on its own. However, the complexity increases significantly when considering the wide range of software engineering tasks that must be addressed. To make progress in developing fully morescient GAI approaches, what is needed is a clear representation of the various dynamic semantics, as well as an understanding of how they are represented across training data sets used to train morescient code models and how they are aligned with the syntactic code model.

\subsubsection*{Lack of Scalable Testing Approaches}

In the preceding subsection on data quality, we touched upon general issues related to ensuring compilability and executability of code. Since true behavior of code modules can only be observed through run-time observations (i.e., software testing), suitable sets of tests are required that exercise the behavior of code in a way that accurately captures its behavior.


Current benchmarking approaches rely on a combination of traditional test generation techniques and LLM-based methods, such as synthesis of test inputs or sequences, exemplified by \textsc{EvalPlus} \cite{liu2023is}. These methods focus primarily on improving coverage metrics (e.g., mutation score) through defect detection. However, they have two significant limitations: (1) they lack a systematic and structured approach to test case design, instead relying on the assumption that defect testing can effectively capture run-time behavior; and (2) their scalability is compromised by the use of classic test representation techniques, such as unit test representations and drivers, which are designed for single-code-module testing and fail to scale to ultra-large execution scenarios. Moreover, these test drivers usually do not have the capability to capture the other interesting run-time behavior -- that is the measurement of non-functional properties including performance and execution time.

To address these limitations, there is a need for novel test case representation languages that can effectively discern subtle nuances in run-time behavior. Additionally, scalable test drivers are required to efficiently execute large corpora of executable code through mass-execution, allowing for the collection of comprehensive run-time data.

\subsubsection*{Fragmentation and Superfluity of Benchmarks}

The community has acknowledged the limitations of traditional NLP-based benchmarks, such as the BLEU score \cite{10.1145/3551349.3556903}, which evaluates generated code by comparing it to canonical ground truth solutions based on textual similarity. Popular benchmarks like \textsc{HumanEval} and \textsc{MBPP}, therefore, use unit-test-driven evaluation with the pass@k metric \cite{chen2021evaluating} (i.e., the probability that at least one of the top k generated code samples for a given problem passes all the unit tests).

Benchmark approaches are often applied manually (i.e., through the manual curation of coding problems in terms of desired behavior and tests) and thus are time-consuming and error-prone \cite{liu2023is}, and/or are used together with arguably ``risky'' LLM-based code synthesis technology (i.e., because models trained on synthetic data generated by GAI may eventually collapse \cite{shumailov2024ai}). The proliferation of these factors has resulted in the fragmentation, superfluity and excessive diversity of benchmarks. Some authors have consequently been driven to propose novel, typically small-scale benchmarks that cater to idiosyncratic preferences, often as a means to sidestep the inherent challenges in evaluating their own approaches.
One of the biggest challenges is curating benchmarks at large scale that contain coding problems and descriptions of desired behavior at appropriate abstraction levels, including black-box (e.g., tests) and white-box behavior (e.g., traces). To overcome these challenges, there is a need for large-scale benchmarking approaches that can automatically create new coding problems (i.e., functionality) and corresponding descriptions of run-time behavior.

\subsubsection*{Lack of Suitable Platforms}


A crucial requirement for developing fully morescient LLMs is the availability of robust platforms that can collect, store, and manage vast quantities of high-quality
observation data, encompassing both functional and non-functional aspects. These platforms must be specifically designed to handle ultra-large data sets and provide suitable data structures for training and improving LLMs. To achieve parity between behavior data sets and code data sets, these platforms must effectively address the limitations and challenges outlined earlier, ensuring the resulting data sets are comprehensive, accurate, and large enough for reliable model development.

\subsubsection*{Challenges in Obtaining Observation Data at Large Scale}

Achieving scalable testing poses two key challenges. First, as mentioned above, there is a need for optimized test drivers that can execute potentially large numbers of tests on large numbers of code modules in parallel, without compromising performance. Second, obtaining precise observation data at scale within a given time budget is essential, but comes with its own set of complexities. A significant obstacle to efficient testing is the need for advanced analysis and tracing techniques. Extensive tracing through analysis tools and measurement calipers can provide valuable insights into code behavior, but introduces a substantial overhead in execution time. This can lead to significantly slower code execution due to the added complexity of instrumentation or introspection techniques. To overcome these challenges, improved techniques that balance scalability with precision are needed.

In general, there is a trade-off between the amount of data traced (i.e., the level of precision of the data), and the overhead caused at execution \cite{4815280}. The lower the precision of the traced data, the lower the overhead. To mitigate this overhead, two key developments are essential: (a) the continued advancement of dynamic analysis techniques to improve their efficiency, such as through hybrid approaches combining static and dynamic analysis \cite{ernst2003static}, and (b) a deeper understanding of the aforementioned trade-offs to support more informed design decisions when capturing morescient data at appropriate precision levels for the software engineering tasks at hand.
\section{Capturing Behavioral Semantics} 
\label{sec:stimulus_response_data_structures}

Computer scientists have developed a range of different techniques for describing the semantics of software (i.e., programming languages) at different levels of abstraction, including denotational techniques, where run-time meaning is defined by mappings to well-known mathematical domains, axiomatic or algebraic techniques, where run-time meaning is defined by the cumulative behavior of component actions, and natural language (sometimes called approximate) techniques, where run-time meaning is defined in textual prose (e.g., code documentation) \cite{gunter1992semantics,Hierons:2009:UFS:1459352.1459354}. In practice, the three most important levels of abstraction are ---

\begin{itemize}
    \item \textit{abstract semantics}: which describe the black-box behavior of functional abstractions (e.g., classes or methods) in terms of their externally visibly behavior (i.e., properties),
    \item \textit{operational semantics}: which describe the white-box behavior of the implementation of a functional abstraction in terms of the individual, step-wise operations that realize its overall behavior (e.g., the method or class body),
    \item \textit{concrete semantics}: which describe the detailed execution semantics of a compiled method or class body at the level of the machine code or intermediate code of the execution platform.
\end{itemize}

To achieve full morescience, GAI code models will ultimately need to be trained on descriptions at all levels and in all the different forms in which they are available. Languages used to capture observation-based descriptions of abstract semantics are currently the most inadequate, as they tend to focus on unit testing rather than data analysis or GAI training. The primary obstacle in addressing these challenges is the lack of suitable data structures that can systematically and scalably capture the abstract semantics of functional abstractions in terms of their actual responses to valid stimuli.

\subsection{Data Structures}

To address this obstacle we propose three new data structures -- \textbf{sequence sheets}, which capture sequences of stimulus-response interactions (cf. tests in terms of triples of inputs, operation invocation and corresponding outputs \cite{6963470}) in a tabular form, \textbf{stimulus-response matrices} (SRMs), which capture collections of sequence sheets representing multiple tests of multiple implementations of a given functional abstraction (i.e., functionality or coding problem), and \textbf{stimulus-response hypercubes} (SRHs) which capture collections of SRMs representing multiple repetitions of multiple tests of multiple functional abstractions.

\begin{figure*}
	\centering
	\includegraphics[scale=0.55,trim=0cm 0cm 9cm 0cm,clip]{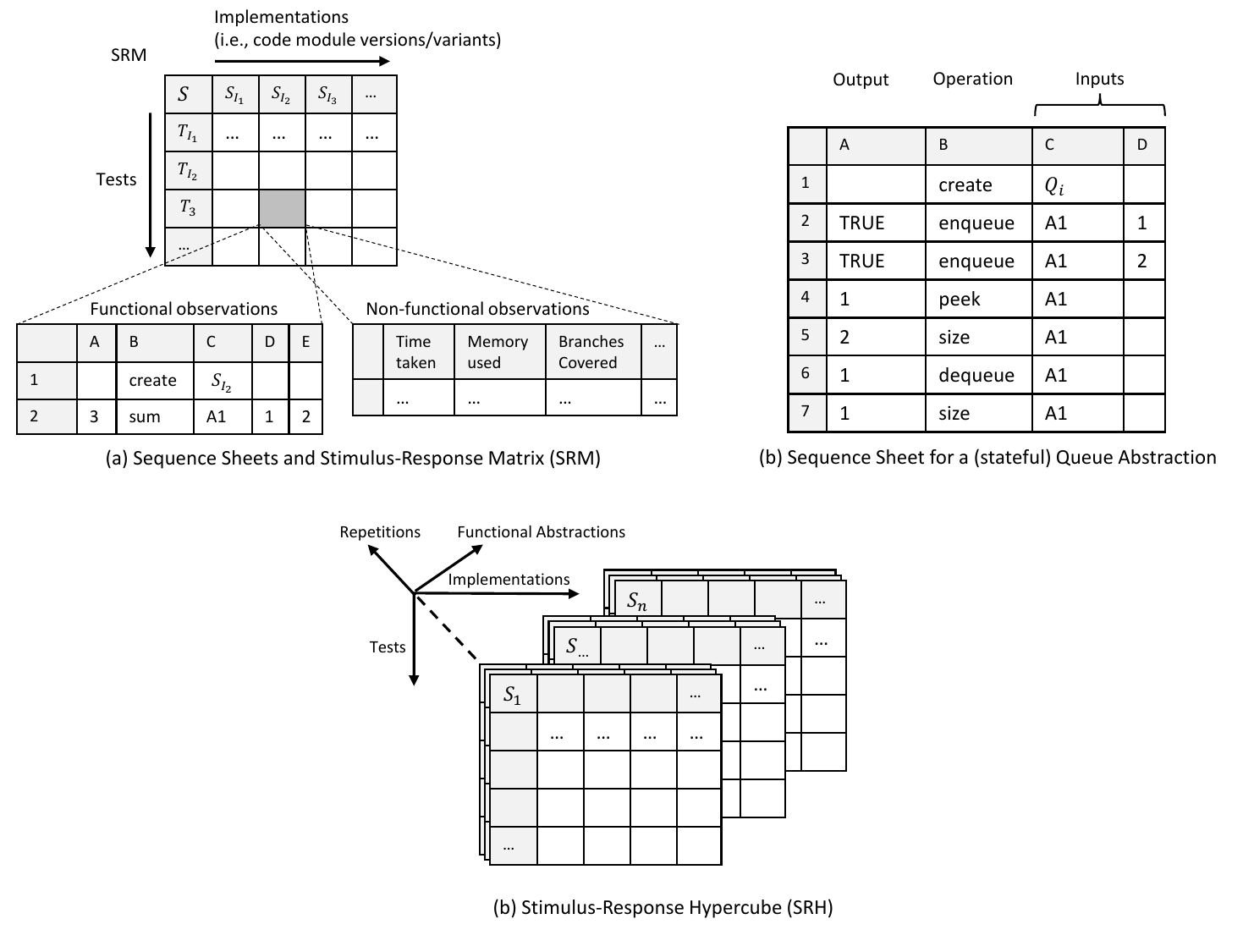}
	\caption{Proposed Data Structures for Behavior Representation}
	\label{fig:srm} 
\end{figure*}

Figure \ref{fig:srm} (a) provides a schematic representation of the first two of these. The bottom left-hand side shows a sequence sheet that captures the response of an implementation of a code module $S$, $S_{I_2}$, to a test $S_{T_3}$, which entails one invocation of the functionality provided by $S$ (a single ``sum'' operation for adding two integers). This corresponds to the single (grey) cell of the SRM, which stores multiple tests of multiple implementations of $S$. As well as the functional information stored in a sequence sheet, each cell also provides access to relevant non-functional information about the test, such as execution time, memory usage and trace information (here branches covered). 

Figure (b) presents a more complex sequence sheet for a (stateful) queue abstraction. Unlike the ``sum'' abstraction, which comprises a single statement, the queue test sequence showcases a series of statements to test core queue behavior -- specifically, the first-in-first-out (FIFO) behavior. Notably, all intermediate observational states are captured (i.e., execution flow), providing a comprehensive view of an implementation's execution over time. SRMs can universally store or link observational records of arbitrary semantic types, encompassing both abstract and operational semantics.

Complex inputs and outputs, such as stateful objects, can be stored in sequence sheets using multiple approaches. The choice depends on the structure of the object. For objects where properties (i.e., class attributes) can be represented by a simple key-value hierarchy, their entire state can be serialized into a suitable representation such as a JSON document, where all properties are accessible by key. Alternatively, when only specific data is needed, inspector methods can be identified that access relevant properties (i.e., state data) of interest. For binary data inputs such as input streams, several strategies exist: either the storage location of the physical copy can be directly linked into the sequence sheet, or suitable metadata like hash signatures can be employed to identify and reference equivalent or similar data. This latter approach reduces storage redundancy while enabling efficient comparison and retrieval of identical or near-identical data.

Finally, Figure (c) gives a schematic representation of an SRH, the third data structure, which collects multiple SRMs applied to multiple code module implementations multiple times in potentially different conditions (e.g., execution environments). SRHs, therefore, provide a multidimensional way of navigating over and analyzing observations from many executions of many implementations under many controlled conditions, including different target execution environments in which observations are obtainable. Since they are representable in the ubiquitous data frame structures supported by popular data processing languages like Python and R, they also lend themselves to analysis by mainstream data analytics tools (i.e., are highly accessible and offer interoperability).

\subsection{Platform and Test Driver}


The key to obtaining more sophisticated data sets containing (1) executable code and (2) run-time data, at scale, lies in automation. The roadmap envisions a platform that not only realizes the data structures outlined above, but also offers flexible, domain-specific languages to design automatic, data-driven workflows on morescient data sets (e.g., SRHs) based on individual selection criteria tailored to the software engineering task at hand. Our prototype implementation of such a platform, the Large-Scale Software Observatorium (\textsc{LASSO})\footnote{project is freely available on GitHub: \url{https://softwareobservatorium.github.io/}} \cite{KESSEL2024111971}, features a domain-specific scripting language to design analysis pipelines for generating high-quality data sets based on individual preferences and functional/non-functional properties of interest \cite{8930881}. It achieves this by integrating a test driver and additional measurement tools to facilitate the mass-execution of code modules and record their run-time functional and non-functional behavior in SRMs. \textsc{LASSO} employs a dedicated ``arena'' test driver that efficiently executes large sets of tests and code module implementations (i.e., specified in stimulus matrices) harvested from repositories or synthesized from LLMs, in a distributed manner (using vertical and horizontal parallelization).
\section{Continually Evolving, Open Stimulus-Response Hypercubes}
\label{sec:continual_srh}

To achieve data quantities for morescient run-time data comparable to those of syntactic code datasets and realize the desired scaling effects of LLMs, we propose developing an ever-growing morescient dataset grounded in the SRH concept introduced earlier. We refer to this concept as an open, continual SRH, which we will describe in the remainder of this section.

\subsection{Dimensional Extensibility}

\begin{figure}
    \centering
    \includegraphics[scale=0.42,trim=0cm 1cm 0cm 0cm,clip]{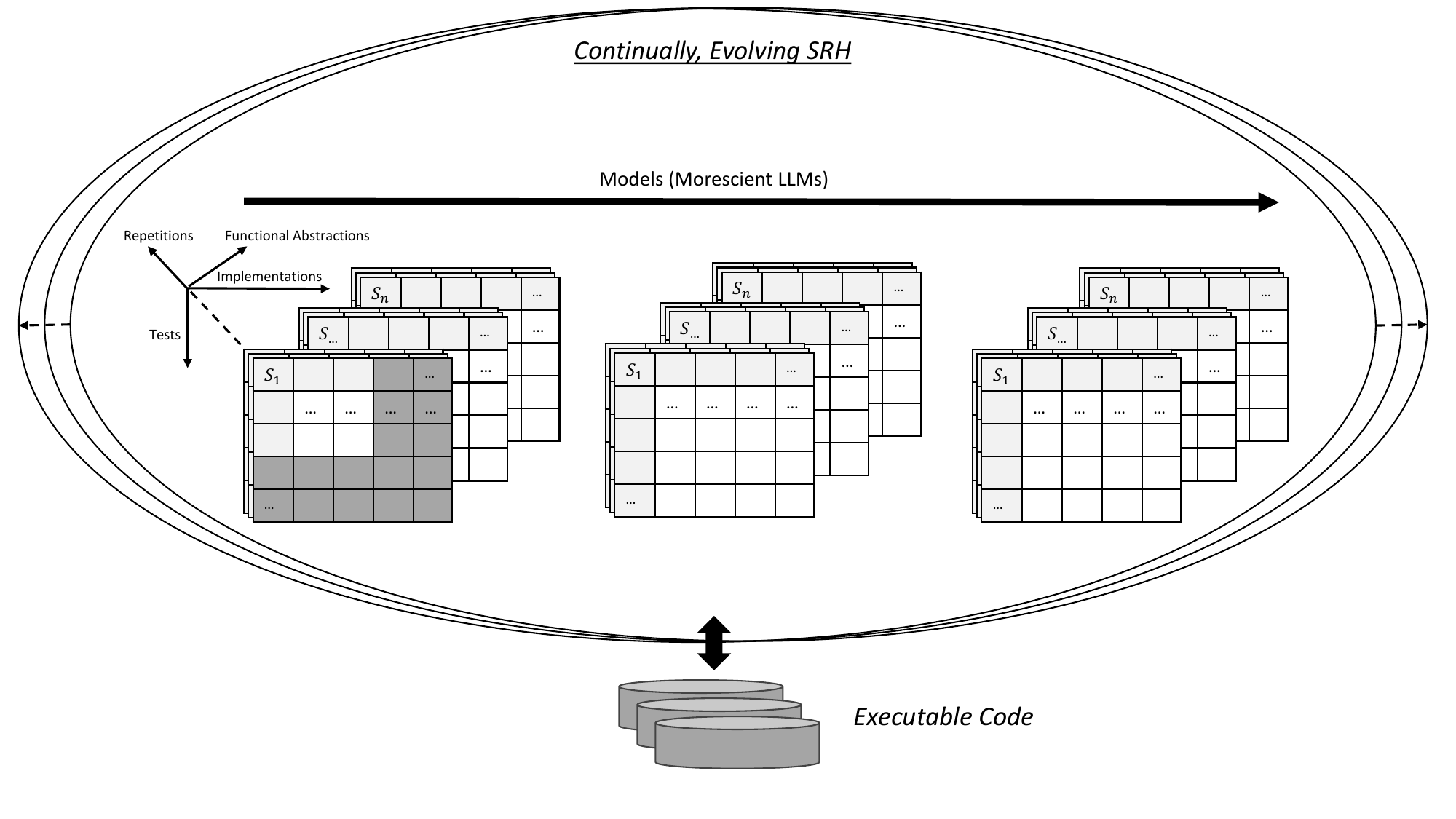}
    \caption{An Open, Continually Evolving SRH of de facto Behavior Data}
    \label{fig:continual_srh}
\end{figure}

There are many potential approaches to developing a continually evolving SRH. However, to ensure its trustworthiness from the outset, we believe it must be designed to be open and accessible. This aligns with the principles of open science and aims to ``democratize'' access to the data, similar to current open-scientific collaborations such as the Big Project and its open code dataset, \textsc{The Stack} \cite{lozhkov2024starcoder2stackv2}. Such an open, community-driven SRH of the kind visualized in Figure \ref{fig:continual_srh} would be continuously expanded and enriched by many contributors over time, including researchers and industry practitioners alike.

At the beginning, such an SRH would be a sparse data structure. The subsequent expansion \cite{KESSEL2024111971} would not only continually refine the core dimensions of the SRH (cf. Section \ref{sec:stimulus_response_data_structures}), incorporating additional harvested/synthesized tests and implementations for each functional abstraction over time, but would also grow to accommodate new dimensions. New data sources (i.e., code repositories), tools and GAI models could be used to harvest or synthesize these. Such a growing SRH would also facilitate the future integration of diverse, behavior data sets by community members using their own morescient GAI approaches in the models dimension. Over time, we anticipate that the community will converge on a common, effective representation of run-time behavior for morescient GAI, agreeing upon them across various levels of abstraction (abstract and operational semantics). This shared understanding will serve as a foundation for further advancements around the open, evolving SRH.

Similar to today's state-of-the-art code LLMs, an evolving SRH must support polyglot programming. This capability will allow the community to leverage a wealth of knowledge from diverse codebases written in various relevant programming languages. To accelerate the scaling effects of morescient LLMs, the community must focus on developing common and effective representations of run-time behavior. To maintain a clear separation of concerns and ensure interoperability, we assume that run-time data and syntactic data (i.e., executable code) will be kept in separate datasets. Thus, SRHs will essentially maintain a link to code module implementations. Since tests are represented as sequence sheets, they are also explicitly stored in SRHs.

With a diverse community helping to evolving SRHs, it is reasonable to assume that the overall quality of the data will improve. As more users engage with the system, they are likely to identify discrepancies or inaccuracies in the data. This collective scrutiny can lead to the detection and correction of errors, ultimately resulting in a more reliable and trustworthy morescient data set (including executable code data sets).

\subsection{Unified Benchmarking}

The open, evolving SRH concept described above not only provides a continuously growing dataset of training data, but also offers a vast repository of sample cases for unified benchmarking. Given that benchmarks are specific to software engineering tasks, individual benchmarks can be derived from the SRH, as long as standard practices in machine learning training like train-validation-test splits are followed to prevent data leakage.

As such an SRH continuously evolves, new revisions (versions) of the data can be created to facilitate ongoing benchmarking with fresh data. This means that not only is the SRH itself expanding and evolving, but the benchmarks used to assess emerging morescient GAI approaches are evolving as well. Consequently, the results shown by current initiatives such as leaderboards can be directly linked to the specific versions of morescient data that produced those results.
\section{A Roadmap for Morescient GAI}
\label{sec:usage}

Sanitized stimulus-response data, stored in the aforementioned data structures, can facilitate morescient GAI in several ways. Below, we outline a roadmap for the next few years, highlighting key milestones and research directions for leveraging observational data to drive progress toward fully morescient GAI approaches.

\subsection{Curating Data Sets}

Initially, the key enabling factor for morescient GAI approaches will be the availability of large, high-quality data sets. A foundational version of an open, evolving SRH described in the previous section will therefore be created early on. More specifically, to obtain high-quality datasets for morescient GAI, it is essential to first focus on developing high-quality, syntax-driven code models that are executable and testable. This foundation is crucial, as morescient GAI relies on these models to capture run-time observations of functional and non-functional properties in terms of the SRHs discussed before. Therefore, the quality of the syntactic code models will directly impact the accuracy of the morescient data curation process. Thus, the creation of morescient data sets will not only complement the syntactic models that currently exist, it will also further improve their quality since non-executable (non-testable) code can be identified and filtered out.

Scalable code analysis platforms \cite{6606588}, such as software observatories (e.g., \textsc{LASSO}) that realize the aforementioned data structures are essential for enabling the mass-execution of code modules at ultra-large scales, thereby generating the vast quantities of run-time data needed to grow the open SRH. Furthermore, automated build systems such as continuous integration platforms can provide a supplementary source of run-time data by continuously building and running code. Despite its potential value, the role and quality of collected build data remain understudied \cite{7962393} and warrant further research to fully realize their exploitation. SRHs can be generated from the build and run-time data collected by these platforms, thereby enhancing the value of what would otherwise be discarded build data. By repurposing this data, SRHs offer a means to preserve the knowledge gained.

\subsection{Training}

Once a foundational, evolving SRH has been established, the first and most crucial step is to include it in LLMs' training data, either during the initial pre-training stage when new LLMs are created or in subsequent transfer learning steps, where existing models are further trained on SRHs to ``fine-tune'' their behavioral awareness. In both scenarios, the observational data encapsulated within SRHs can be made available for training in either a direct or indirect form. In direct approaches, LLMs are presented with ``raw'' (cleaned) SRH data containing uncensored descriptions of numerous executions of a wide array of code modules.
 
While this is likely to give the best results, the scale of the data sets involved presents numerous practical problems to researchers, including the need for significant processing power. In indirect approaches, the SRH data is (pre)processed in some way as preparation for training. For example, reusable vector representations (e.g., embeddings such as \textsc{Word2vec} \cite{mikolov2013efficient} and \textsc{BERT} \cite{devlin-etal-2019-bert}), can be generated from the data in which information with ``close'' vector values is assumed to be more closely related than that with more distant values. Another indirect approach is to construct knowledge graphs from SRHs in order to encode relationships efficiently and to train models to predict missing links \cite{10.1145/3424672}.

State-of-the-art architectures used to train LLMs have achieved impressive performance on natural language texts, including code, leveraging their syntactic capabilities. However, incorporating observational data alongside syntactic data as training inputs may necessitate architectural modifications. The choice of representation for observational data has a significant impact on the construction of more sophisticated models that integrate both facets. Observational data can be treated either as textual content or as an additional modality, similar to image information. Given our assumption that both syntactic and observational data will co-evolve, combining models in the spirit of multi-modal architectures may yield benefits. With the availability of diverse run-time data, it may be advantageous to specify multiple representations of observational data, encompassing various levels of abstraction and semantic precision (Section \ref{sec:limitations}). Given the preliminary findings on observational data presented in Section \ref{sec:limitations}, we anticipate that further research will focus on developing suitable architectures and representations for training morescient code models.

\subsection{Augmented Generation}

The second way is to use an evolving SRH, along with a test driver like \textsc{LASSO}'s arena, as an external knowledge source to provide a variety of services that existing, syntactically-trained code models can call to improve their accuracy and relevance. These generation augmentation services \cite{mialon2023augmented} can range from simple testing services, where the LLM can ask whether a generated piece of code has the expected functionality, to oracle services, where the LLM can ask for oracle values for the desired functionality (see \cite{kesselGAI2024} for further details). The external knowledge source therefore essentially provides an external fact-checking mechanism that can identify ``incorrect'' code generations. Recently, researchers have been exploring fact-checking models \cite{tang2024minicheckefficientfactcheckingllms} that specialize in verifying the output of LLMs, helping to detect false claims and hallucinations. These models leverage factual information to verify whether an LLM's output is factually accurate. Morescient versions of these models can take advantage of knowledge sources like SRHs to validate a claim by comparing it against established facts (here observational data).

\subsection{Prompting}

Prompting LLMs involves designing the input prompts to elicit the most accurate, relevant and helpful responses from a model. The design of the prompts used to invoke LLM models, therefore, plays a huge role in their perceived performance \cite{10440574,LI2024112002}. The aforementioned stimulus-response data structures in Section \ref{sec:stimulus_response_data_structures} can help improve prompting in three main ways. First, the minimalistic and structured test representation approach offered by sequence sheets can reduce user errors and misunderstandings when they serve as prompt templates. Secondly, since tests are frequently included in prompts to code LLMs to improve precision, users can enrich their prompts to non-morescient models using information from SRHs. Thirdly, the service APIs (e.g., functions) of the platforms that support morescient GAI can themselves be included in prompts to encourage LLMs to obtain external factual information \cite{mialon2023augmented}.

\subsection{Test-driven Software Experimentation}

Finally, the adoption and development of any new technology, including morescient GAI, is critically dependent on experimental evidence of its strengths and weaknesses. Stimulus-response data structures coupled with a scalable test driver of the kind offered by \textsc{LASSO}, are a key foundation for evaluating and comparing code LLMs in a generalizable way. Only by performing large scale test-driven experiments can subtle differences in the generation of alternative code solutions or tests be detected and ``n-version'' comparisons of the different implementations be performed (e.g., strengthened through differential testing \cite{10440574}, or differential GAI in general \cite{kesselGAI2024}), for example to fine-tune LLM ``hyperparameters''.

Moreover, making detailed information about the continual SRH openly accessible in accordance with the principles of open science \cite{openinitiate2023} facilitates the development of new and improved evaluation metrics, such as pass@k \cite{chen2021evaluating}, or BLEU scores, which rely on a canonical, ground truth implementation \cite{10.1145/3551349.3556903}.

\subsection{AI-driven Software and Decision-Making}

Orthogonal to the milestones previously described, the integration of generative AI into software products is on the rise, blurring the lines between algorithmic and probabilistic reasoning. Components of a software system can now comprise imperative code and probabilistic models, making it increasingly difficult to distinguish between them. This shift contrasts with retrieval augmented generation, where a model is enhanced by external tools.

AI-driven software tools, such as smart IDEs (e.g., \textsc{VSCode} and GitHub \textsc{Copilot}), can leverage the standardized data structures (i.e., sequence sheets and SRHs) to achieve interoperability between LLM representations and run-time information obtained in IDEs. This will enable seamless use of algorithmic or LLM-based components for software engineering tasks and decision-making. These tools will likely employ multiple models and meta-modeling, making standard data structures and semantic representations even more essential to ensure interoperability between them.

For now, the developer remains actively engaged in the development process. Data structures, particularly sequence sheets, not only serve as a means to understand and communicate behavior but also facilitate collaboration among stakeholders -- like code. However, with advancements in AI-driven software engineering, it is likely that some of the tasks will soon be fully automated. There is growing interest in LLM-based agents like \textsc{SWE-agent} \cite{yang2024sweagentagentcomputerinterfacesenable} and \textsc{AutoDev} \cite{tufano2024autodevautomatedaidrivendevelopment}, which utilize a combination of models and tools to accomplish specific tasks within a given cost budget, thereby achieving complete autonomy. As LLM-based agents evolve, the integration of sophisticated data structures like SRHs out-of-the-box will become even more important, further accelerating the automation of software engineering tasks.

In an ideal future, where morescient GAI has reached its full potential, it may become possible for highly advanced morescient models to accurately predict the execution of software, making traditional verification and validation processes largely obsolete or redundant. With developers and autonomous agents increasingly reliant on data-driven decision-making, an open, evolving SRH would provide a robust foundation for informed decision-making over time (e.g., reduce debugging activities or log analysis).
\section{Conclusion}
\label{sec:conclusion}

The hypothesis of this paper is that the trustworthiness, and thus the utility, of GAI for software engineering tasks will be significantly boosted by creating morescient code models trained on the semantic (i.e., dynamic) as well as the syntactic (i.e., static) aspects of software. Among other things, making LLMs morescient will allow them to make much better judgments about the run-time behavior of the software they synthesize in response to prompts. However, creating, organizing and representing the vast quantities of observation data needed to realize this vision will require new kinds of data structures and software observation platforms that can populate them while respecting important data set properties, such as avoiding duplication \cite{10.1145/3359591.3359735}. 

We have introduced the new data structures and the platform we envisage and have discussed the different ways in which they can help foster morescient GAI in an open, community-driven way. In a seminal ``road map'' for source code analysis published in 2007 \cite{4221615}, Brinkley predicted that by 2025 software analysis tools will ``appear to understand algorithms and can automatically suggest superior solutions to specific problems''. By applying the ideas, technologies and roadmap presented in this paper, we believe Brinkley's vision is well on track to realization, and fully morescient GAI will soon facilitate the practical use of GAI in mainstream software projects.


\bibliographystyle{ACM-Reference-Format}
\bibliography{literature}


\end{document}